\documentclass[lettersize,journal,9pt]{IEEEtran}
\usepackage{subfiles}
\usepackage[T1]{fontenc}    
\usepackage{booktabs}       
\usepackage{amsmath,amsfonts}
\usepackage{bm}
\usepackage{braket}
\usepackage{graphicx}
\usepackage{latexsym}
\usepackage{comment}
\usepackage[hang,small,bf]{caption}
\usepackage[subrefformat=parens]{subcaption}

\newcommand{\fig}[1]{Fig.~\ref{#1}}
\newcommand{\secref}[1]{Sec.~\ref{#1}}
\newcommand{\Eqref}[1]{Eq.~\eqref{#1}}

\newcommand{\lc}{ \left\{ } 
\newcommand{\rc}{ \right\} } 
\newcommand{\lp}{ \left( } 
\newcommand{\rp}{ \right) } 
\newcommand{\CBra}[1]{ \lc #1 \rc } 
\newcommand{\Par}[1]{ \lp #1 \rp } 

\newcommand{\Ceil}[1]{ \left\lceil #1 \right\rceil }

\newcommand{\inv}[1]{ \frac{1}{#1} }
\newcommand{\ZO}{\CBra{0, 1}}

\newcommand{\binlog}[1]{\log_{2}{#1}}

\newcommand{\ctext}[1]{\raise0.2ex\hbox{\textcircled{\scriptsize{#1}}}}

\newcommand{\EstQ}[1]{\left\langle{#1}\right\rangle}

\newcommand{\im}{\mathit{i}}

\newcommand{\order}[1]{ \mathit{O}\Par{ #1 } }

\newcommand{\pbeta}{\bm{\beta}}
\newcommand{\pgam}{\bm{\gamma}}

\newcommand{\pz}{\bm{z}}

\newcommand{\paramState}[1]{ \Ket{\psi \Par{ #1 }} }
\newcommand{\paramStateT}[1]{ \Bra{\psi \Par{ #1 }} }

\newcommand{\numCounter}{ M }

\newcommand{\txTime}[1]{ t_{ \text{#1} } }
\newcommand{\ResetT}{ \txTime{reset} }
\newcommand{\InitT}{ \txTime{init} }
\newcommand{\LayerT}{ \txTime{L} }
\newcommand{\MeasureT}{ \txTime{meas} }
\newcommand{\CircuitT}{ \txTime{QC} }

\newcommand{\BW}[1]{ BW_{ \text{#1} } }
\newcommand{\BWMeas}{ \BW{meas} }
\newcommand{\BWInst}{ \BW{inst} }
\newcommand{\BWProposed}{ \BW{proposed} }
\newcommand{\BWMSB}{ \BW{MSB} }
\newcommand{\BWnonMSB}{ \BW{non-MSB} }

\newcommand{\rzT}{ t_{R_z} }

\newcommand{\entT}{ t_{Ent} }
\newcommand{\rxT}{ t_{R_x} }

\begin{document}
%
\title{Inter-temperature Bandwidth Reduction\\in Cryogenic QAOA Machines}

\author{Yosuke Ueno, Yuna Tomida, Teruo Tanimoto, Masamitsu Tanaka, Yutaka Tabuchi, Koji Inoue, Hiroshi Nakamura\vspace{-7mm}
\thanks{Y. Ueno and Y. Tabuchi are with RIKEN, Y. Tomida and H. Nakamura are with the University of Tokyo, T. Tanimoto and K. Inoue are with Kyushu University, and M. Tanaka is with Nagoya University.}
}


\maketitle



\begin{abstract}
  The bandwidth limit between cryogenic and room-temperature environments is a critical bottleneck in superconducting noisy intermediate-scale quantum computers. This paper presents the first trial of algorithm-aware system-level optimization to solve this issue by targeting the quantum approximate optimization algorithm. Our counter-based cryogenic architecture using single-flux quantum logic shows exponential bandwidth reduction and decreases heat inflow and peripheral power consumption of inter-temperature cables, which contributes to the scalability of superconducting quantum computers.
\end{abstract}

\IEEEpeerreviewmaketitle

\section{Introduction}
\begin{figure}[t]
  \centering
  \includegraphics[width=\columnwidth]{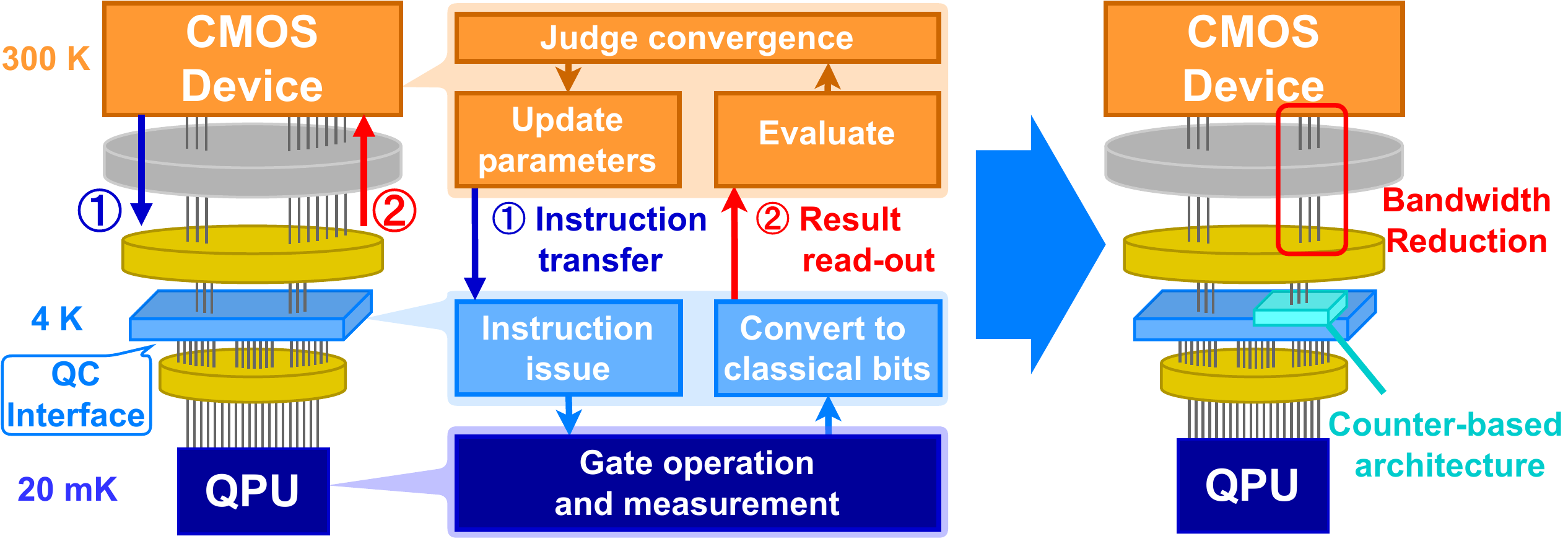}
  \vspace{-4.5mm}
  \caption{Baseline system design of SQC (left), QAOA procedure (middle), and our proposed system design (right).}
  \label{fig:cryo-qaoa}
\end{figure}

Quantum computers (QCs) require classical computers (CCs) for their essential control and management to perform quantum computation. Examples of such interactions between quantum devices and CCs include quantum error correction in fault-tolerant quantum computing and classical-quantum hybrid algorithms in noisy intermediate-scale quantum (NISQ) machines, such as Variational Quantum Algorithms (VQAs) and quantum approximate optimization algorithm~(QAOA)~\cite{farhi2014qaoa} shown in Fig.~\ref{fig:cryo-qaoa} (middle). 
Consequently, QCs need to ensure sufficient bandwidth between a quantum processing unit~(QPU) and CCs for essential interactions.

Superconducting qubits operate in a dilution refrigerator, as shown on the left side of Fig.~\ref{fig:cryo-qaoa}, due to their susceptibility to thermal noise. 
Superconducting QCs (SQCs) use high-frequency coaxial cables for inter-temperature interaction between a room-temperature CC and a cryogenic QPU. 
The heat inflow from these cables strictly constrains the system volume of SQCs. 
As the SQC systems scale up, the number of required cables for the interaction also grows, leading to a higher heat inflow through the cables and more power consumption for the peripherals of cables.
In addition to the cooling cost required for the QPU itself, this increased heat inflow and peripheral power consumption may exceed the cooling capacity of the refrigerator. 
In such a case, the thermal noise induces decoherence in the QPU, leading to incorrect computations.
Thus, the bandwidth is a critical system parameter as it defines the number of cables or their material.

We propose a concept to compress information within the cryogenic environment to reduce the required inter-temperature bandwidth in SQCs during the NISQ algorithms such as VQAs.
To show the effectiveness of our concept, we focus on QAOA, one of the simplest variants of VQAs, as a first step. We analyze the communication behavior of QAOA and propose a novel counter-based cryogenic architecture for it. 
Our contributions are summarized as follows.

\begin{enumerate}
\item We analyze the inter-temperature communication of QAOA to identify the bandwidth bottleneck (\secref{analysis}).
\item We propose a new counter-based architecture to reduce the inter-temperature bandwidth and design it with single-flux quantum~(SFQ) circuits (\secref{proposal}). 
\item Our evaluation shows that the bandwidth requirement is reduced from $ \order{N} $ to $\order{1}$ by using $\order{\log{N}}$-bit counters where $ N $ is the number of qubits (\secref{evaluation:tradeoff}).
\item We discuss the trade-off between the power dissipation of cables and our counter-based architecture (\secref{evaluation:scalability}).
\end{enumerate}

\section{Background}
\label{background}

\subsection{Quantum approximate optimization algorithm}
\label{background:qaoa}
QAOA\cite{farhi2014qaoa} is the leading example of VQAs for combinatorial optimization in which the optimization is reduced to the search for ground states of the Hamiltonian of the Ising model
\begin{align}
  H_C =
  - \sum_{i=1}^N h_i Z_i
  - \sum_{i \ne j} J_{ij} Z_i Z_j
  \label{eq:hamiltonian},
\end{align}
where $ Z_i $ represents the spin orientation and $ h_i, J_{ij} $ represents the magnitude of the magnetic field and interaction between spins\cite{inagaki2016coherent}.
The output state of the QAOA circuit is formulated as
\begin{align}
  \Ket{\psi(\pgam, \pbeta)}
  = e^{ - \im \beta H_B }
  \Par{ \prod_{l=1}^{L} U_B(\beta_l) U_C(\gamma_l) }
  \Ket{+}^{\otimes N} ,
\end{align}
where $U_C(\gamma) = e^{ - \im \gamma H_C }, U_B(\beta) = e^{ - \im \beta H_B }$, and $H_B = - \sum_{i=1}^N X_i $. Note that $ \pgam, \pbeta $ are parameters to optimize.

A significant challenge in the use of QAOA lies in appropriately mapping classical optimization problems into a quantum framework.
Mapping such a problem onto the QAOA necessitates a translation process where the objective function and constraints of the optimization problem are reformulated into the Ising model. 
Intuitively, we can imagine this process as representing each integer decision variable by a set of qubits.
For example, in a max cut problem, each spin $Z_i$ in Eq.~(\ref{eq:hamiltonian}) indicates the group to which the $i$-th node of the target graph belongs.
See Ref.~\cite{xie2022ising} for more details on mapping combinatorial optimization problems to the Ising model.

\subsection{Sampling nature of cost function in QAOA}
\label{background:sampling}
In QAOA, we minimize the expected value of energy
$
  \EstQ{H_C} = \paramStateT{\pgam,\pbeta} H_c \paramState{\pgam,\pbeta}.
$
Since it is impossible to measure this energy directly, we approximate it with the cost function
\begin{align}
  C(\pz) = 
  \sum_{i=1}^{N} s_{i}z_{i} +
  \sum_{i \ne j} c_{ij} \Par { z_{i} \oplus z_{j} },
  \label{eq:cost}
\end{align}
which receives the classical bit sequence $ \pz \in \ZO^{N} $ obtained from the quantum state measurement. Note that $ s_i $ and $c_{ij} $ correspond to $ h_i $ and $J_{ij} $ in Ising Hamiltonian.
We take $ T $ times of sampling (trials) and then approximate the energy as
\begin{align}
  \EstQ{H_C} \simeq \inv{T} \sum_{t=1}^T C(\pz_t).
  \label{eq:sampling}
\end{align}

\subsection{Related work on inter-temperature bandwidth reduction of SQCs}
Using cryogenic computing peripherals such as SFQ circuits for peripherals of SQCs has been actively studied\cite{jokar2022digiq, opremcak2021measurement}.
Jokar \textit{et al.}~\cite{jokar2022digiq} proposed an SFQ-based qubit control system based on the SIMD concept. 
While their system focuses on the bandwidth for qubit control, our system mainly concentrates on the measurement readouts based on the bottleneck analysis on \secref{analysis}.

In the field of physics, Opremcak \textit{et al.}~\cite{opremcak2021measurement} proposed a method for efficient qubit measurements in cryogenic environments by equipping photon detectors on an SFQ chip. 
Based on such a cryogenic qubit measurement technique, this study presents a system architecture for reducing inter-temperature communications for qubit measurements.

Moreover, to the best of our knowledge, this is the first paper that discusses such algorithm-aware system-level optimization for NISQ machines.
Unlike the prior works above~\cite{jokar2022digiq,opremcak2021measurement}, we concentrate on application-aware optimization as a new research direction.

\section{Modeling Communication on QAOA machines}
\label{analysis}

\subsection{Baseline system design}
\begin{figure}[t]
  \centering
  \includegraphics[width=0.95\columnwidth]{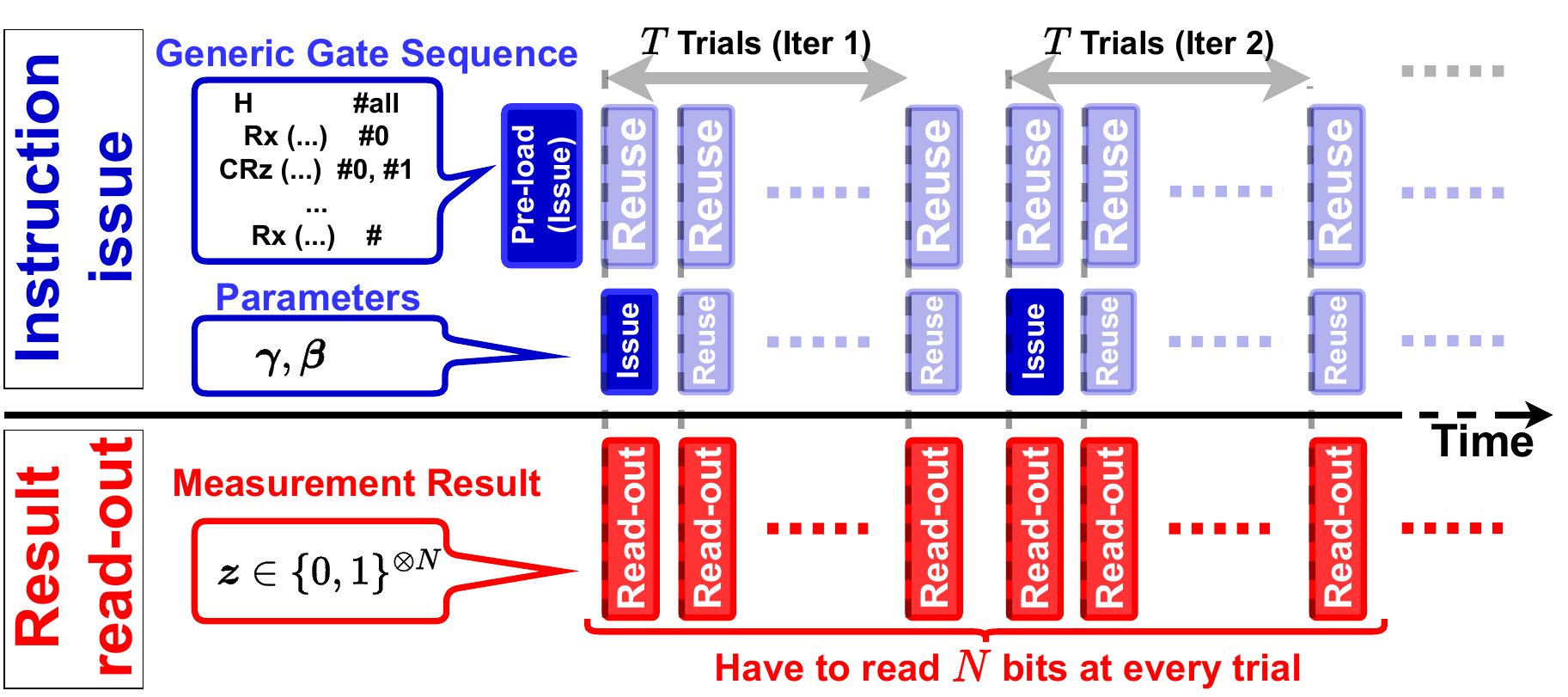}
  \vspace{-1mm}
  \caption{Timing of inter-temperature communication in QAOA}
  \label{fig:comm-timing}
  \vspace{-2mm}
\end{figure}

The left side of Fig.~\ref{fig:cryo-qaoa} shows the baseline system. We assume that SFQ pulses implement QPU control signals and the inter-temperature communication is digital. 
The execution flow of QAOA is a repetition of the cycle shown in the center of \fig{fig:cryo-qaoa}. Inter-temperature communications occur during \ctext{1} the instruction transfer for QPU control and \ctext{2} the read-out of measurement results as shown in \fig{fig:comm-timing}.
In the following subsections, we model the required bandwidths of these communications on the baseline system and discuss their major bottleneck.

\subsection{Bandwidth for inter-temperature communication}
\label{analysis:bandwidth}
First, we formulate the execution time of the entire QAOA circuit $t_{QC}$ to model the required bandwidths.
The quantum circuit of QAOA repeats the same sequence of gates (which we call a \textit{layer}) differing only in parameters following the Hamiltonian of \Eqref{eq:hamiltonian} derived from the problem instance. Specifically, for the first- and second-order terms $S, C$ of the Hamiltonian, each layer consists of $ S$ $R_z $ gates, $ C $ entangling blocks ($Ent$ := CNOT-$R_z$-CNOT), and $ N $ $ R_x $ gates. 

We generally denote the time required to apply a quantum gate $ G $ as $t_{G}$. $\ResetT$, $\InitT$, and $\MeasureT$ represent the reset, initialization, and measurement operation time per qubit, respectively.
We assume the gate level parallelism of one-qubit gates, including measurement, as $P$ and that of two-qubit gates as $P/2$. 
Then, we model the per-layer execution time $\LayerT$ and $t_{QC}$ as 
\begin{align}
\LayerT &= S \rzT + 2C \entT + N \rxT,\label{eq:exec-time-layer}\\ 
t_{QC} &=
  \cfrac{
    N \ResetT 
    +
    N \InitT 
    +
    L \LayerT
    +
    N \MeasureT 
  }{ P },~
  \label{eq:exec-time-bare}
\end{align}
where $L$ represents the number of layers per QAOA circuit.

\noindent
\textbf{\ctext{1} Bandwidth for instruction transfer: }
We analyze the locality of instruction transfer communication. Each layer is the same sequence of gates differing only in parameters. Therefore, we need to transfer instructions of a layer only once and can reduce the total number of instruction transfers by treating the generic gate sequence and QAOA parameters separately, as shown in \fig{fig:comm-timing}.
The layer is consistent throughout the optimization cycle, so it suffices to transfer the layer instructions only once at a low speed before QAOA execution.
The following discussion does not consider the layer instruction transfer for the bandwidth analysis because it is short enough relative to $t_{QC}$. The parameters must only be transferred once per circuit parameter update, \textit{i.e.}, once every $T$ trials. Such a buffering functionality is plausible for SFQ circuits by using ring buffers\cite{ishida2020pipelinedsfq}.

Based on the discussion above, we model the required bandwidth for the instruction transfer $\BWInst$.
Let $b_p$ be the bit width of each QAOA parameter. The transfer of $ 2 l b_p $ bits of QAOA parameters must be completed before the $ l $-th layer operation starts in the first trial, \textit{i.e.}, within
$ \left(N (\ResetT + \InitT) + (l-1) \LayerT\right) / P$ in each iteration. The maximum bandwidth requirement is 
\begin{align}
  \BWInst = \underset{l \in \CBra{1, \cdots, L}}{\text{ max }}
  \frac{ 2 P l b_p }{ N \Par{ \ResetT + \InitT } + (l-1) \LayerT } \notag \\
  = \max \CBra{
    \frac{2P b_p}{ N \Par{ \ResetT + \InitT }},
    \frac{2P L b_p}{ (L-1) \LayerT }
  }.
  \label{eq:inst-bw}
\end{align}

\noindent
\textbf{\ctext{2} Bandwidth for measurement read-out: }
The required bandwidth for read-out $\BWMeas$ can be estimated as $ N $ bits per quantum circuit execution, as shown in the of \fig{fig:diagram-counter},
\textit{i.e.},
\begin{align}
  \BWMeas = N / t_{QC} = 
  \frac{ P }{ \ResetT + \InitT + L \LayerT/N + \MeasureT }.
  \label{eq:measure-bw}
\end{align}

\subsection{Attention to measurement bandwidth at design phase}
\label{analysis:design-time-bandwidth}
When designing the system, the resources should be sufficient to handle the different problems and parameter settings of the algorithm. Thus, we consider the worst case in terms of bandwidth, \textit{i.e.}, the shortest execution time case.

Equations~(\ref{eq:inst-bw}) and~(\ref{eq:measure-bw}) both have $\LayerT$ in the denominator, which depends on $ S $ and $ C $ and represents the sparseness of the problem. For example, for a maximum cut problem, $ C $ corresponds to the number of edges in the graph. In the worst case, these values are $ S = 0, C = N-1 $ because, for the case with $ C < N-1 $, there exists a qubit that does not entangle, and we can split the problem into smaller problems that do not require $N$ qubits. In this case, $t_L$ in Eq.~(\ref{eq:exec-time-layer}) comes to $O(N)$. 
Also, we assume the number of layers to be $ L = 1 $ for the shortest execution time. 
Hence, in the worst case, the orders of $\BWInst$ and $\BWMeas$ are $O(P b_p / N)$ and $O(P)$, respectively.
Given that $b_p$, the bit width of the parameters to be optimized, is at most several tens of bits, it results in the inequality $ P b_p / N \ll P \iff b_p / N \ll 1 $ as the number of qubits $ N $ scales to $ 10^2 $ or larger. 
Based on these observations, we conclude that the $\BWMeas$ is dominant in the inter-temperature bandwidth of the baseline, and the following sections focus on reducing it.

\section{Counter-based architecture at the 4-K stage}
\label{proposal}

\subsection{Bandwidth reduction with counter-based architecture}
\label{proposal:counting}
\begin{figure}[t]
  \centering
  \includegraphics[width=0.95\columnwidth]{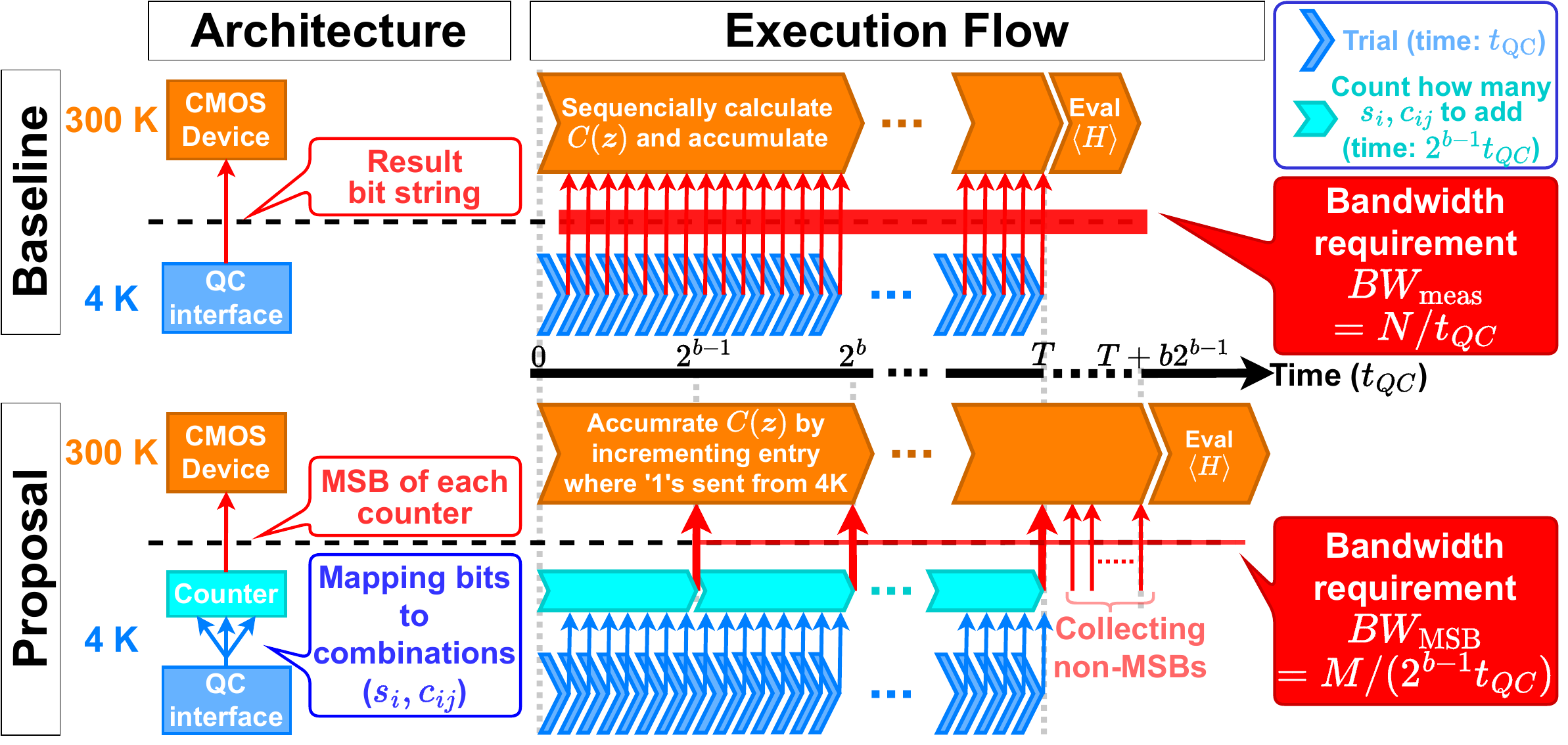}
  \vspace{-2mm}
  \caption{Inter-temperature communication in QAOA using counters}
  \vspace{-2mm}
  \label{fig:diagram-counter}
\end{figure}

In the baseline system, we calculate $\EstQ{ H_C }$ in Eq.~(\ref{eq:sampling}) by evaluating $C(\pz)$ in Eq.~(\ref{eq:cost}) at every trial. Thus, we need inter-temperature communications to send measurement results of $N$ qubits to the room-temperature environment after each trial ends, as shown in \fig{fig:diagram-counter}. The required bandwidth of the baseline system is $\BWMeas = N/t_{QC}$ as explained in Section~\ref{analysis}.

By contrast, our counter-based system keeps counting values of qubit measurements in a 4-K environment throughout $ T $ trials to reduce the inter-temperature communication.
We count the number of measurements of `1's in each qubit and the non-zero coefficients in each qubit pair as 
$C_{i} = \sum_{t=1}^{T} z_{i}^{t}$ and $C_{ij} = \sum_{t=1}^{T} \Par{ z_{i}^{t} \oplus z_{j}^{t} }$, respectively, throughout $ T $ trials to calculate the cost function as 
\begin{align}
  \EstQ{H_C} \simeq
  \inv{T} \Par{
    \sum_{i=1}^{N} s_{i}C_{i} +
    \sum_{i\ne j} c_{ij}C_{ij}
  }.
\end{align}
In other words, the function $\EstQ{H_C}$ is estimated in the sum of products using the counter values after $ T $ trials.
Assuming that each counter has a sufficient bit width to keep $C_{i}$ and $C_{ij}$ throughout $T$ trials, our method requires only one inter-temperature communication with $ \numCounter \Ceil{ \binlog{T} }$bits after all $T$ trials. 
Here, $ \numCounter $ represents the number of counters in use. 

However, applying our method naively could increase the bandwidth requirement rather than the baseline because transferring all the counter values at once may require a larger bandwidth than $\BWMeas$. 
In addition, we cannot assume that the counter bit width is always sufficient as the trial count $ T $ changes depending on use cases.

To avoid these problems, we introduce an MSB-sending policy to level the bit transfers across all $ T $ trials and suppress the bandwidth requirement. 
In this policy, we limit each counter entry to $ b $ bits ($ < \Ceil{ \binlog{T}} $), clear~(destructively read) MSBs once every $2^{b-1}$ trials, and then transfer them to the upper stage. 
As shown in \fig{fig:diagram-counter}, the CMOS device in the upper stage takes these MSBs as ``$ 2^{b-1} $ counts'' and stores them as the upper bits of counters.
Hence, the CMOS and SFQ devices cooperatively behave as large-bit-width counters. 
While this policy requires more than $ \numCounter \Ceil{ \binlog{T} }$ bits to transfer, it reduces the bandwidth requirement by smoothing out the communication across all $T$ trials rather than transferring all bits at once.

In contrast to the bandwidth of the baseline system $\BWMeas = N/\CircuitT$, in our system, $ \numCounter $bits of MSBs are transferred once per $2^{b-1}$ trials. Thus, the bandwidth requirement for MSB $\BWMSB$ is just $ \BWMSB = \numCounter / (2^{b-1}\CircuitT)$. Hence, the bandwidth reduction is exponential to the counter entry $b$.

Note that we can further reduce this bandwidth requirement by adaptively transferring MSBs only when one or more entries are filled with all `1's. However, we use the regular MSB-sending policy to simplify the circuit design in this paper.

\subsection{Execution time overhead for collecting non-MSBs}
\label{proposal:tradeoff}
In our method, we collect $ b \numCounter $ bits remaining on the SFQ counters after all $ T $ trials to complete the cost evaluation. 
Suppose that we collect these bits within $ t_{C} $. The bandwidth requirement for the non-MSBs collecting $\BWnonMSB$ equals $b \numCounter / t_{C} $. Hence, the bandwidth of our system, $\BWProposed$, is 
\begin{align}
\BWProposed = 
\max{
  \Par {
    \BWMSB, \hspace{3pt}
    \BWnonMSB
  }
}.
\end{align}
To keep $\BWnonMSB$ as small as $\BWMSB$, $ t_C $ must be long enough to satisfy $ t_C \ge b 2^{b-1} \CircuitT $. 
Hence, our method has a trade-off between the bandwidth reduction and the execution time overhead; the bandwidth reduction is exponential to $ b $ while the execution time grows from $ T \CircuitT $ to $ (T+ b 2^{b-1}) \CircuitT $.

\subsection{Configuration of counters with SFQ circuit}
\label{proposal:counter}
\begin{figure}[t]
  \centering
  \includegraphics[width=0.98\columnwidth]{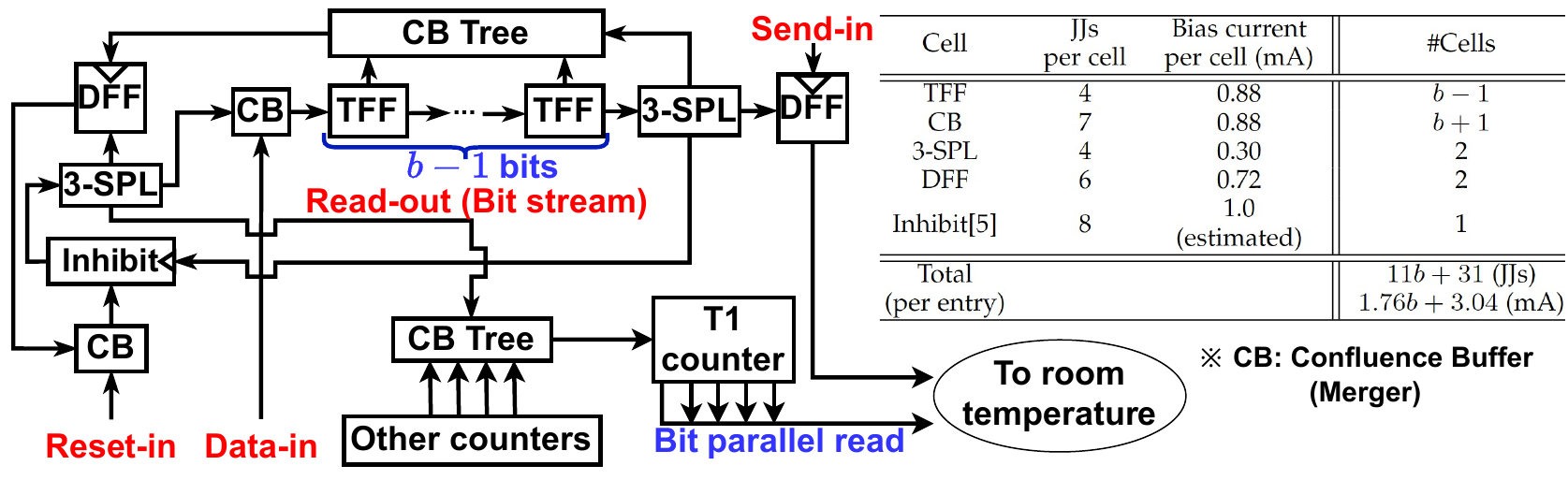}
  \vspace{-1mm}
  \caption{Detailed configuration of a counter entry}
  \vspace{-2mm}
  \label{fig:counter}
\end{figure}

We designed our counter-based architecture using the existing RSFQ cell library~\cite{detail_of_cell_library_ADP2}, and \fig{fig:counter} shows the circuit diagram, the number of JJs, and the total bias current per counter entry. Note that we estimated the bias current of the \textit{inhibit} cell\footnote{The inhibit cell has two inputs: an inhibiting signal and a data signal. If the data signal arrives before the inhibiting signal, the data passes through the gate; otherwise, the gate output is `0'~\cite{tzimpragos2020temporal}.} from the number of JJs because it is not presented in Ref.~\cite{tzimpragos2020temporal}.

The counter module consists of $N(N+1)/2$ counters.
Each bit consists of a T-flipflop (TFF) except for the MSB, which uses a D-flipflop (DFF).
During the execution of the quantum circuit, pulses are routed round-robin to the $N(N+1)/2$ ``send'' lines to achieve communication smoothing in the MSB-sending policy.

In the non-MSB collection phase, the reset signal is routed to each counter, and a bit stream is generated on the read-out line. Assuming the counter has a value $ v $, this bit stream consists of $2^{b}-v$ `1's and stops when a pulse is routed to the inhibit cell~\cite{tzimpragos2020temporal} after the counter overflows.
This bit stream has $ \order{2^b} $ length, and sending it as is increases the bandwidth.
Therefore, we first route it to a \textit{T1~counter}\cite{polonsky1994bff}, which allows bit-parallel read, placed after the CB tree and shared among all counter entries.
Here, we can perform this routing by triggering each counter sequentially; thus, only one T1 counter is sufficient. Our design requires only one reset signal per counter, while clock and read-out trees are needed to use bit-parallel readable T1 counters for all counter entries.

\section{Evaluation}
\label{evaluation}

\subsection{Trade-off between bandwidth and execution time}
\label{evaluation:tradeoff}

\begin{figure}[t]
    \centering
    \includegraphics[width=0.95\columnwidth]{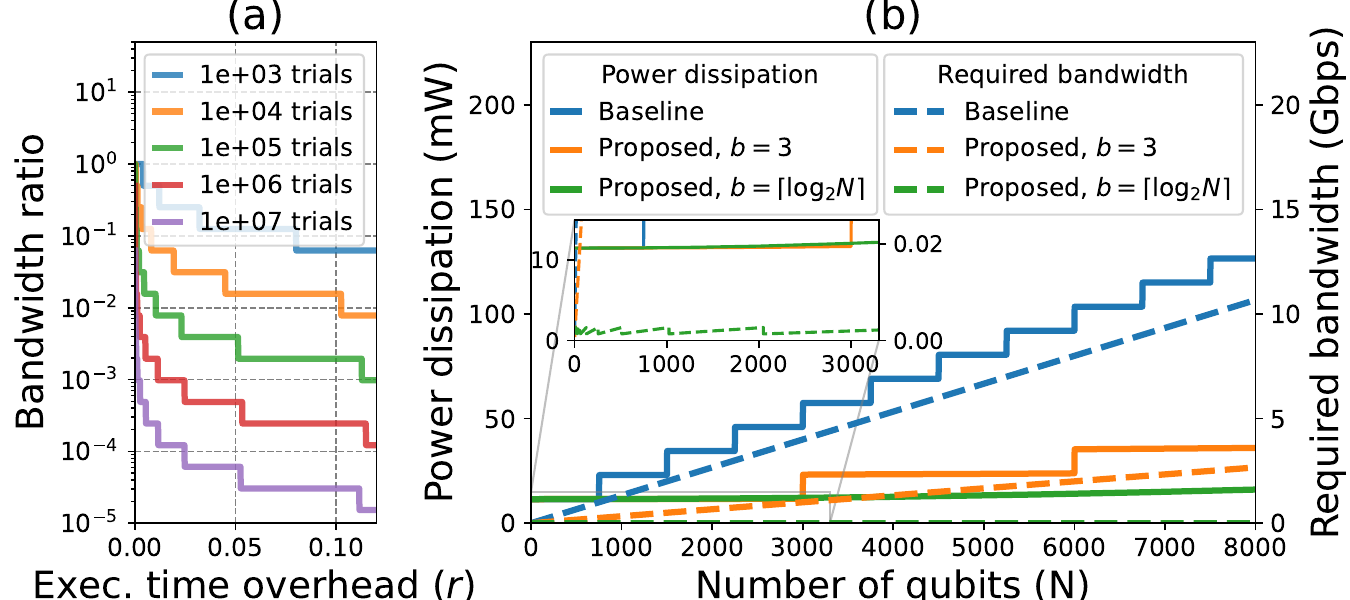}
  \vspace{-1mm}
  \caption{(a) Bandwidth reduction to execution time overhead and (b)~power dissipations of baseline and proposed systems.}
  \label{fig:eval}
  \vspace{-4mm}
\end{figure}

As discussed in \secref{proposal:tradeoff}, the execution time of QAOA in our proposed system gets $ 1 + b 2^{b-1} / T $ times longer than that of the baseline system, Here, we study how large we can set $ b $ within an acceptable execution time overhead $r$.
In other words, we evaluate how much bandwidth we can reduce with an increased execution time by a factor of $ 1+r $. The counter bit width $ b $ must satisfy
\begin{align}
  \frac{ b 2^{b-1} }{ T } < r
  \hspace{2mm} \Leftrightarrow \hspace{2mm}
  b 2^b < 2 r T.
  \label{eq:bound}
\end{align}
As discussed in \secref{analysis:bandwidth}, we consider the worst case~($ \numCounter = N-1 $). The bandwidth ratio between the proposed and baseline is $\cfrac{\BWProposed}{\BWMeas} = \cfrac{\numCounter/2^{b-1} t_{QC}}{N t_{QC}} \simeq \cfrac{1}{2^{b-1}} $. Figure~\ref{fig:eval}~(a) shows the bandwidth ratio achieved by maximizing $ b $ under \Eqref{eq:bound} at a certain value of $r$. Note that each plot of \fig{fig:eval}~(a) has a staircase shape because the number of bits in the counter $b$ is an integer.
Our proposed system works well for larger $T$. For example, when the acceptable execution time overhead is 5\% ($r = 0.05$), the proposed system reduces the bandwidth by $99.993$\% with $ T = 10^{7} $ and $ b = 15 $, or $87.5$\% ($b = 4$) with $ T = 10^{3} $ and $ b = 4 $.

$ T $ is the number of samplings to approximate the quantum state and note that it is still unknown how many samplings are required for QAOA to find a good solution.
Here, all the $T$ values shown in \fig{fig:eval}~(a) are within $ \order{N}$ to $\order{N^3}$ as we assume NISQ machines with qubits on the order of $ 10^2$ to $10^3 $.

\subsection{Impact on the scalability of the number of qubits\label{evaluation:scalability}}
First, we show the bandwidth reduction achieved by the SFQ counters with fixed or $O(\log{N})$-size bit lengths. 
As discussed in \secref{analysis:design-time-bandwidth}, the bandwidth requirement of the baseline system is $ \order{P}$, \textit{i.e.}, proportional to the gate parallelism. 
On the other hand, that of our system is $\order{P/2^b}$ as demonstrated in \secref{evaluation:tradeoff} with a certain time overhead $r$ according to Eq.~(\ref{eq:bound}).
Considering the worst case for the bandwidth with $P = N$ (fully parallelized) and using parameters described in references \cite{jokar2022digiq,opremcak2021measurement} as $\ResetT = 100$~ns, $\InitT = t_{R_x} = t_{R_z} = 10$~ns, $t_{CNOT} = 60$~ns, and $\MeasureT = 380$~ns, respectively, $t_{QC}$ in Eq.~(\ref{eq:exec-time-bare}) is estimated as $t_{QC} \simeq 100 \times 1 + 10 \times 3 + 60 \times 4 + 380 \times 1 = 750$~ns. 
Using the $t_{QC}$ value, Fig.~\ref{fig:eval}~(b) shows the required bandwidths of the baseline and our system ($b=3$) as blue and orange dashed lines, respectively.
Furthermore, our system can reduce the required bandwidth to $ \order{ 1 } $ by setting $ b = \Ceil{\log{ N }}$ (green dashed line in the figure, see the enlarged view). 
Note that we assume $\BWMSB$ is dominant for the proposed system's bandwidth in this evaluation, \textit{i.e.}, Fig.~\ref{fig:eval}~(b) does not care about $r$ in Sec.~\ref{evaluation:tradeoff}.

Next, we evaluate the trade-off between heat inflow through cables and power consumption in a cryogenic environment with our system. 
We suppose that the systems have stainless steel coaxial cables (UT085 SS-SS) between the 4-K and upper stages, and its heat inflow is 1.0~mW as in Tab.~2 of Ref.~\cite{krinner2019cryosetup}. 
In addition to the heat inflow, we consider the power consumption of the cable peripherals.
We assume the amplifier, which is the main source of peripheral power consumption, is LNF-LNC4\_8C as in Ref.~\cite{krinner2019cryosetup}, and its power consumption is 10.5~mW per cable according to its datasheet. 
The bandwidth per cable is assumed to be 1~Gbps, and the system has a sufficient number of cables to exceed its required bandwidth (\textit{e.g.}, the system whose bandwidth requirement is 2.5~Gbps has three cables).

Based on the RSFQ design in \secref{proposal:counter} and the ERSFQ power model in Ref.~\cite{mukhanov2011energy}, we estimate the power consumption of our system when ERSFQ is applied.
Here, the power of the counter, $P_{\text{counter}}$, with ERSFQ can be estimated as 
$P_{\text{counter}} =(\text{bias current}) \times (\text{frequency}) \times \Phi_{0} \times 2$.
We use flux quantum $\Phi_0$ of $2.068$~fWb, and the counter is assumed to operate at 1.33~MHz, the reciprocal of $t_{QC} = 750$~ns.
As a result, $P_{\text{counter}}$ is $(9.71b+16.8)$~pW.

Table~\ref{tab:cable_counter_configuration} summarizes the configuration of cables and our SFQ counter, and Fig.~\ref{fig:eval}~(b) compares the power dissipations (sum of heat inflow and power consumption) between the baseline and proposed systems.
While that of the baseline is dominated by cables as the bandwidth increases, the proposed system reduces the impact of cables by reducing the bandwidth with the negligible overhead of counters. 
Our system achieved lower power dissipation than the baseline on QAOA machines with more than 750 qubits.

Note that our method focuses on cables between 300~K and 4~K environments and does not change the number of cables between 4~K
and 20~mK. 
Consequently, the power consumption of our architecture remains as in Tab.~\ref{tab:cable_counter_configuration}, without causing additional heat transfer.

The essential advantages of our method are the exponential bandwidth reduction and power consumption of the order of pW.
While the evaluation of this paper utilizes a particular cable and peripheral, our method is expected to work well across a variety of system designs because of its general advantages.

\begin{table}[bt]
\centering
\scriptsize
\caption{Configuration of cables and SFQ counters\label{tab:cable_counter_configuration}}
\vspace{-2mm}
\begin{tabular}{|c||l|l|} \hline
                                                        & Power dissipation                                                                                                                        & Configuration                                           \\ \hline \hline
\begin{tabular}[c]{@{}l@{}}Coaxial\\ cable\end{tabular} & \begin{tabular}[c]{@{}l@{}}Heat inflow: 1.0 mW\cite{krinner2019cryosetup}\\ Periphelars: 10.5 mW\cite{krinner2019cryosetup}\end{tabular} & One cable per 1~Gbps                                      \\ \hline
\begin{tabular}[c]{@{}l@{}}SFQ\\ counter\end{tabular}   & $(9.71b+16.8)$ pW                                                                     & \begin{tabular}[c]{@{}l@{}}$M = N(N+1)/2$\\ ERSFQ (Freq. = 1.33~MHz )\end{tabular} \\ \hline
\end{tabular}
\vspace{-2mm}
\end{table}

\section{Conclusion and future work}
We proposed a general concept of information compression within the cryogenic environment to reduce inter-temperature communication for NISQ machines. 
To show its effectiveness, we targeted QAOA as a first step.
We modeled the inter-temperature communication in QAOA and proposed a bandwidth-efficient counter architecture with SFQ circuits.
Our method reduces the communication by transferring the MSB of each counter that tallies the number of additions of each cost function term throughout trials.
It successfully reduced the required bandwidth exponentially and decreased heat inflow and peripheral power consumption of coaxial cables in a cryogenic environment with the negligible overhead of SFQ counters.

Computations involving general VQAs often require a vast array of information compared to QAOA.
Our concept of information compression using counters is expected to work effectively even in general VQAs; however, a straightforward implementation of a counter architecture capable of concurrently storing all the information required for general VQAs is inefficient and not feasible from a hardware standpoint. 
Our future work is to propose practical strategies, such as sampling a subset of the information with counters, and apply them to the proposed architecture to support general VQAs.

\ifCLASSOPTIONcompsoc
  \section*{Acknowledgments}
\else
  \section*{Acknowledgment}
\fi

This work was partly supported by JST PRESTO Grant Number JPMJPR2015, JST Moonshot R\&D Grant Number JPMJMS2067, JSPS KAKENHI Grant Numbers JP22H05000, JP22K17868, RIKEN Special Postdoctoral Researcher Program.



\bibliographystyle{IEEEtran}
\bibliography{IEEEabrv,ref-en}

\begin{thebibliography}{10}
\providecommand{\url}[1]{#1}
\csname url@samestyle\endcsname
\providecommand{\newblock}{\relax}
\providecommand{\bibinfo}[2]{#2}
\providecommand{\BIBentrySTDinterwordspacing}{\spaceskip=0pt\relax}
\providecommand{\BIBentryALTinterwordstretchfactor}{4}
\providecommand{\BIBentryALTinterwordspacing}{\spaceskip=\fontdimen2\font plus
\BIBentryALTinterwordstretchfactor\fontdimen3\font minus
  \fontdimen4\font\relax}
\providecommand{\BIBforeignlanguage}[2]{{%
\expandafter\ifx\csname l@#1\endcsname\relax
\typeout{** WARNING: IEEEtran.bst: No hyphenation pattern has been}%
\typeout{** loaded for the language `#1'. Using the pattern for}%
\typeout{** the default language instead.}%
\else
\language=\csname l@#1\endcsname
\fi
#2}}
\providecommand{\BIBdecl}{\relax}
\BIBdecl

\bibitem{farhi2014qaoa}
\BIBentryALTinterwordspacing
E.~Farhi \emph{et~al.}, ``{A Quantum Approximate Optimization Algorithm},''
  2014. [Online]. Available: \url{https://arxiv.org/abs/1411.4028}
\BIBentrySTDinterwordspacing

\bibitem{inagaki2016coherent}
T.~Inagaki \emph{et~al.}, ``{A coherent Ising machine for 2000-node
  optimization problems},'' \emph{Science}, vol. 354, no. 6312, pp. 603--606,
  2016.

\bibitem{xie2022ising}
S.~Xie \emph{et~al.}, ``{Ising-CIM: A Reconfigurable and Scalable Compute
  Within Memory Analog Ising Accelerator for Solving Combinatorial Optimization
  Problems},'' \emph{IEEE J. Solid-State Circuits}, vol.~57, no.~11, pp.
  3453--3465, 2022.

\bibitem{jokar2022digiq}
M.~R. Jokar \emph{et~al.}, ``{DigiQ: A Scalable Digital Controller for Quantum
  Computers Using {SFQ} Logic},'' in \emph{Proc. of HPCA}, 2022, pp. 400--414.

\bibitem{opremcak2021measurement}
A.~Opremcak \emph{et~al.}, ``{High-Fidelity Measurement of a Superconducting
  Qubit Using an On-Chip Microwave Photon Counter},'' \emph{Phys. Rev. X},
  vol.~11, p. 011027, Feb 2021.

\bibitem{ishida2020pipelinedsfq}
K.~Ishida \emph{et~al.}, ``32 {GHz} 6.5 {mW} gate-level-pipelined 4-bit
  processor using superconductor single-flux-quantum logic,'' in \emph{Proc. of
  Symp. VLSI Circuits}, 2020, pp. 1--2.

\bibitem{detail_of_cell_library_ADP2}
Y.~Yamanashi \emph{et~al.}, ``100 {GHz} demonstrations based on the
  single-flux-quantum cell library for the 10 {kA}/cm$^2$ {Nb} multi-layer
  process,'' \emph{IEICE Trans. Electron.}, vol.~93, no.~4, pp. 440--444, 2010.

\bibitem{tzimpragos2020temporal}
G.~Tzimpragos \emph{et~al.}, ``{A Computational Temporal Logic for
  Superconducting Accelerators},'' in \emph{Proc. of ASPLOS}, 2020, p.
  435–448.

\bibitem{polonsky1994bff}
S.~Polonsky \emph{et~al.}, ``{Single flux, quantum B flip-flop and its possible
  applications},'' \emph{IEEE Trans. Appl. Supercond.}, vol.~4, no.~1, pp.
  9--18, 1994.

\bibitem{krinner2019cryosetup}
S.~Krinner \emph{et~al.}, ``{Engineering cryogenic setups for 100-qubit scale
  superconducting circuit systems},'' \emph{EPJ Quantum Technol.}, vol.~6,
  no.~1, p.~2, 2019.

\bibitem{mukhanov2011energy}
O.~A. {Mukhanov}, ``{Energy-Efficient Single Flux Quantum Technology},''
  \emph{IEEE Trans. Appl. Supercond.}, vol.~21, no.~3, pp. 760--769, 2011.

\end{thebibliography}
%
%
%

\end{document}